\documentclass[aps,twocolumn,superscriptaddress,prl]{revtex4}
\usepackage{amssymb}
\usepackage{amsmath}
\usepackage{graphicx}

\begin{document}
\title{Hyperfine-mediated gate-driven electron spin resonance}
\author{E.\ A.\ Laird\footnote[1]{These authors contributed equally to this work.}}
\affiliation{Department of Physics, Harvard University, Cambridge, Massachusetts
02138, USA}
\author{C.\ Barthel\footnotemark[1]}
\affiliation{Department of Physics, Harvard University, Cambridge, Massachusetts
02138, USA}
\author{E.\ I.\ Rashba}
\affiliation{Department of Physics, Harvard University, Cambridge, Massachusetts
02138, USA}
\affiliation{Center for Nanoscale Systems, Harvard University, Cambridge, Massachusetts
02138, USA}
\author{C.\ M.\ Marcus}
\affiliation{Department of Physics, Harvard University, Cambridge, Massachusetts
02138, USA}
\author{M.\ P.\ Hanson}
\affiliation{Materials Department, University of California at Santa Barbara,
Santa Barbara, California 93106, USA}
\author{A.\ C.\ Gossard}
\affiliation{Materials Department, University of California at Santa Barbara,
Santa Barbara, California 93106, USA}

\begin{abstract}
An all-electrical spin resonance effect in a GaAs few-electron double quantum dot is investigated experimentally and theoretically.  The magnetic field dependence and absence of associated Rabi oscillations are consistent with a novel hyperfine mechanism.  The resonant frequency is sensitive to the instantaneous hyperfine effective field, and the effect can be used to detect and create sizable nuclear polarizations.  A device incorporating a micromagnet exhibits a magnetic field difference between dots, allowing electrons in either dot to be addressed selectively.
\end{abstract}

\maketitle 

The proposed use of confined electron spins as solid-state qubits~\cite{LossDiVincenzo} has stimulated progress in their manipulation and detection~\cite{JelezkoESR, RugarESR, ElzermanSingleShot, XiaoESR, JohnsonT1, PettaT2, KoppensESR}.  In such a proposal, the most general single-qubit operation is a spin rotation.  One technique for performing arbitrary spin rotations is electron spin resonance (ESR)~\cite{EngelESR}, in which a pair of magnetic fields is applied, one static (denoted~$\mathbf{B}$) and one resonant with the electron precession (Larmor) frequency  (denoted~$\mathbf{\tilde{B}}$).  Observing single-spin ESR is challenging because of the difficulty of combining sufficient~$\mathbf{\tilde{B}}$ with single-spin detection~\cite{JelezkoESR, RugarESR, XiaoESR}. In GaAs quantum dots, where a high degree of spin control has been achieved~\cite{ElzermanSingleShot, JohnsonT1, PettaT2}, ESR was recently demonstrated using a microstripline to generate $\mathbf{\tilde{B}}$~\cite{KoppensESR}.

An alternative to ESR is electric dipole spin resonance (EDSR)~\cite{BellEDSR, McCombeEDSR, RashbaShekaBook}, in which an oscillating \emph{electric} field $\mathbf{\tilde{E}}$ replaces $\mathbf{\tilde{B}}$.  EDSR has the advantage that high-frequency electric fields are often easier to apply and localize than magnetic fields, but requires an interaction between $\mathbf{\tilde{E}}$ and the electron spin.  Mechanisms of EDSR include spin-orbit coupling and inhomogeneous Zeeman coupling~\cite{RashbaShekaBook, KatoGTMR, GolovachLoss,TokuraSlanting}.

In this Letter, we present the first experimental study of a novel EDSR effect mediated by the random inhomogeneity of the nuclear spin orientation.  The effect is observed via spin-blocked transitions in a few-electron GaAs double quantum dot.  For $B=|\mathbf{B}|<1$~T the resonance strength is independent of $B$ and shows no Rabi oscillations as a function of time, consistent with a theoretical model we develop but in contrast to other EDSR mechanisms.  We make use of the resonance to create nuclear polarization, which we interpret as the backaction of EDSR on the nuclei~\cite{GueronRyter, DobersNMR, KoppensESR, BaughPolarization}.  Finally, we demonstrate that spins may be individually addressed in each dot by creating a local field gradient.

\begin{figure}[b!]
\vspace{-0.3 cm} \centering \label{fig:fig1}
\includegraphics[width=3in]{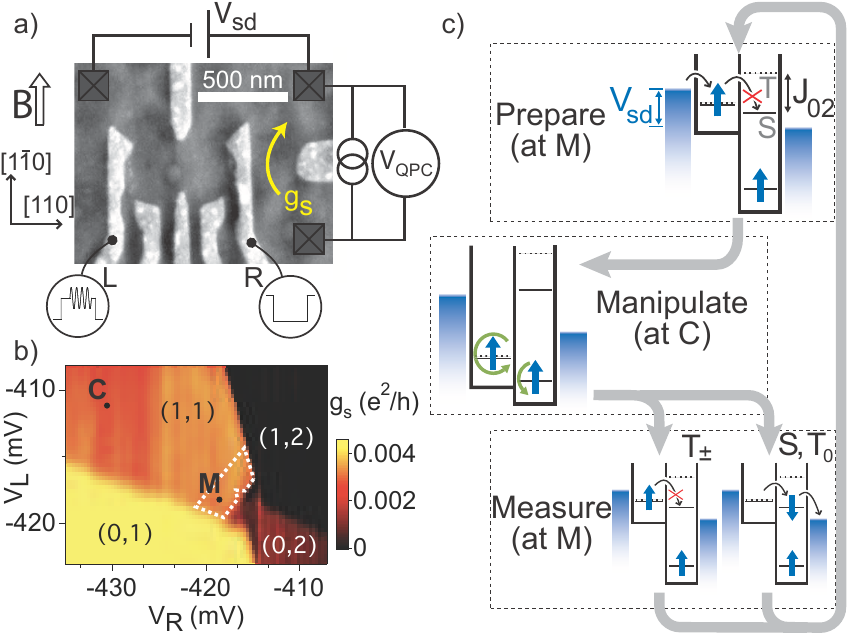}
\vspace{-0.3 cm} \caption{\footnotesize{(Color online) (a)
Micrograph of a device lithographically identical to the one measured, with schematic of the measurement circuit.  The direction of $\mathbf{B}$ and the crystal axes are indicated.  (b) QPC conductivity $g_\mathrm{s}$ measured at $V_\mathrm{sd}\sim 600\mu\mathrm{eV}$ near the (1,1)-(0,2) transition. The spin blockade region is outlined.  Equilibrium occupations for different gate voltages are shown, as are gate configurations during the measurement/reinitialization (M)  and manipulation (C) pulses.  A plane background has been subtracted.  (c) Energy levels of the double dot during the pulse cycle (See text).}}
\end{figure}

The device for which most data is presented (Fig.~1(a)) was fabricated on a GaAs/Al$_\mathrm{0.3}$Ga$_\mathrm{0.7}$As heterostructure with two-dimensional electron gas (density $2\times10^{15}$~m$^{-2}$, mobility $20$ m$^2$/Vs)  110 nm below the surface. Ti/Au top gates define a few-electron double quantum dot.  A charge sensing quantum point contact (QPC), tuned to conductance  $g_\mathrm{s} \sim 0.2e^2/h$, is sensitive to the electron occupation $(N_\mathrm{L},N_\mathrm{R})$ of the left and right dots \cite{FieldSensing, ElzermanSensing}.  The voltages $V_\mathrm{L}$ and $V_\mathrm{R}$ on gates L and R, which control the equilibrium occupation, are pulsed using a Tektronix AWG520; in addition, L is coupled  to a Wiltron 6779B microwave source gated by the AWG520 marker.  A static in-plane field $\mathbf{B}$ was applied parallel to $[1\overline{1}0]$. Measurements were performed in a dilution refrigerator at 150~mK electron temperature, known from Coulomb blockade width.  

As in previous measurements \cite{KoppensESR}, we detect spin transitions with the device configured in the spin blockade regime~\cite{OnoSpinBlockade, JohnsonSpinBlockade}.  In this regime, accessed by tuning $V_\mathrm{L}$ and $V_\mathrm{R}$, a bias $V_\mathrm{sd}$ across the device induces transport via the sequence of charge transitions $(0,2)\rightarrow(0,1)\rightarrow(1,1)\rightarrow(0,2)$.  Intra-dot exchange interaction $J_{02}$ makes the $(1,1)\rightarrow(0,2)$ transition selective in the two-electron spin state, inhibited for the $m_s=\pm1$ triplets $T_\pm$ but allowed for the $m_s=0$ triplet $T_0$ or singlet $S$.  Since decay of $T_\pm$ requires spin relaxation, it becomes the rate-limiting step in transport, and so the time-averaged occupation is dominated by the (1,1) portion of the transport sequence. Figure~1(b) shows the conductance $g_\mathrm{s}$ of the  charge sensor as a function of $V_\mathrm{L}$ and $V_\mathrm{R}$.  Inside the outlined region, where spin blockade is active (see~\cite{JohnsonSpinBlockade} for details), $g_\mathrm{s}$ has the value corresponding to (1,1).

EDSR is detected via changes in $g_\mathrm{s}$ while the following cycle of gate pulses~\cite{KoppensESR} is applied to $V_\mathrm{L}$ and $V_\mathrm{R}$ (Fig.~1(c)). Beginning inside the spin blockade region (M in Fig.~1(b)) initializes the two-electron state to $(1,1)T_\pm$ with high probability.  A $\sim$1~$\mu$s pulse to point C prevents electron tunneling regardless of spin state.  Towards the end of this pulse, a microwave burst of duration $\tau_\mathrm{EDSR}$ at frequency~$f$ is applied to gate L.  Finally the system is brought back to M for $\sim$3~$\mu$s for readout/reinitialization.  If and only if a spin (on either dot) was flipped during the pulse, the transition $(1,1)\rightarrow(0,2)$ occurs, leading to a change in average occupation and in $g_\mathrm{s}$.  If this transition occurs, subsequent electron transitions reinitialize the state to $(1,1)T_\pm$ by the end of this step, after which the pulse cycle is repeated.  This pulsed EDSR scheme has the advantage of separating spin manipulation from readout.

\begin{figure}[t!]
\center \label{fig:fig2}
\includegraphics[width=3in]{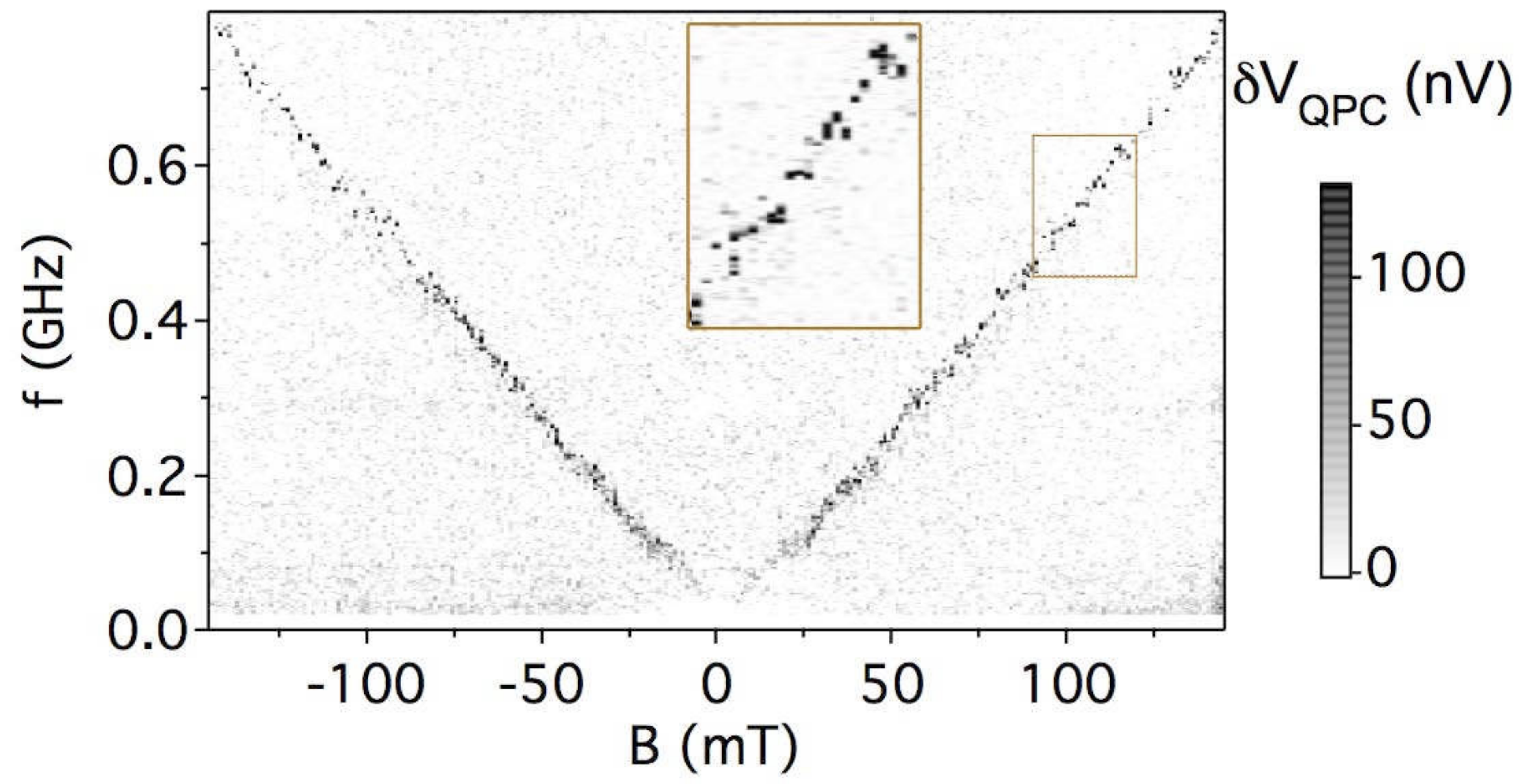}
\vspace{-0.3 cm} \caption{\footnotesize{Signal of spin resonance $\delta V_\mathrm{QPC}$ as a function of magnetic field $B$ and microwave frequency $f$.  EDSR induces a breaking of spin blockade, which appears as a peak in the voltage across the charge sensor $\delta V_\mathrm{QPC}$ at the Larmor frequency.  Field- and frequency-independent backgrounds have been subtracted.  Inset: Jitter of resonant frequency due to Overhauser shifts.}}
\end{figure}

Changes in $g_\mathrm{s}$ are monitored via the voltage $V_\mathrm{QPC}$ across the QPC sensor biased at 5~nA.  For increased sensitivity, the microwaves are chopped at 227~Hz and $\delta V_\mathrm{QPC}$ is synchronously detected using a lock-in amplifier with 100 ms time constant.  We interpret $\delta V_\mathrm{QPC}$ as proportional to the spin-flip probability during a microwave burst.

Resonant response is seen clearly as $B$ and $f$ are varied for constant $\tau_\mathrm{EDSR}=1~\mu$s (Fig.~2.)  A peak in $\delta V_\mathrm{QPC}$, corresponding to a spin transition, is seen at a frequency proportional to $B$.  This is the key signature of spin resonance.  From the slope of the resonant line we deduce for the $g$-factor $|g|=0.39\pm0.01$, typical of similar GaAs devices~\cite{GGKondo, HansonZeeman}.  We attribute fluctuations of the resonance frequency (inset of Fig.~2) to instantaneous Overhauser shifts; their range is $\sim\pm22$~MHz, corresponding to a field of $\sim$~4~mT, consistent with Overhauser fields in similar devices~\cite{KoppensNuclei, JohnsonT1, PettaT2}.

\begin{figure}[t!]
\center \label{fig:fig3}
\includegraphics[width=2.6in]{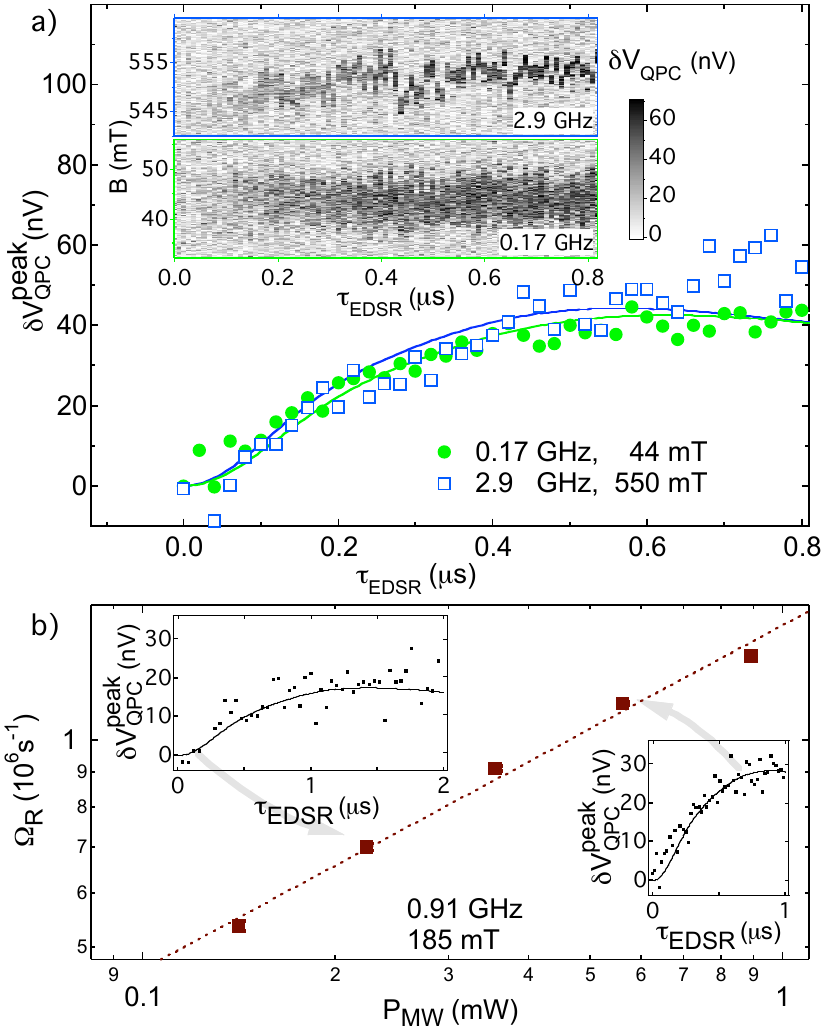}
\vspace{-0.3 cm} \caption{\footnotesize{(Color online) (a) EDSR peak strength $\delta V_\mathrm{QPC}^\mathrm{peak}$ versus microwave pulse duration $\tau_\mathrm{EDSR}$ for two frequencies at equal power, along with theoretical curves.  Inset: Raw data from which the points in the main figure are taken.  Each vertical cut corresponds to one point in the main figure.  Jitter in the field position of the resonance reflects time-dependent Overhauser shifts.  (b) Spin-flip rate $\Omega_R$ as a function of applied microwave power $P_\mathrm{MW}$, along with a fit to the form $\Omega_R \propto \sqrt{P_\mathrm{MW}}$ (dashed line).  Insets: $\delta V_\mathrm{QPC}^\mathrm{peak}$ versus $\tau_\mathrm{EDSR}$ for two values of the microwave power, showing the fits from which points in the main figure are derived.}}
\end{figure}

Behavior of the EDSR peak as a function of duration, strength, and frequency of the microwave pulse is shown in Fig.~3.  To reduce the effects of the shifting Overhauser field, the microwave source is frequency modulated at 3~kHz in a sawtooth pattern with depth 36~MHz about a central frequency $\overline{f}$. Scanning over $B$ for a range of $\tau_\mathrm{EDSR}$ (inset of Fig.~3(a)), the resonance strength $\delta V^\mathrm{peak}_\mathrm{QPC}$ is extracted from a Gaussian fit in $B$.  For $\overline{f}= 0.17$~GHz and 2.9~GHz and equal microwave power~\cite{footnote2}, the results are plotted in Fig.~3(a)   The two curves are similar in turn-on time and saturation value; this is the case for frequencies up to $\overline{f}=6$~GHz.  From similar data (insets of Fig.~3(b)) taken at $\overline{f}=0.91$~GHz, using theory to be described, we extract the dependence of the spin-flip rate $\Omega_R$ on microwave power $P_\mathrm{MW}$ shown in Fig.~3(b). Oscillations in $\delta V^\mathrm{peak}_\mathrm{QPC}(\tau_\mathrm{EDSR})$ are not observed for any $P_\mathrm{MW}$ or $B\lesssim 1$~T.

A theoretical description of $\delta V^\mathrm{peak}_\mathrm{QPC}(\tau_\mathrm{EDSR})$ and its dependence on $B$ and $P_\mathrm{MW}$ can be obtained by modeling EDSR as arising from the coupling of an electron in a single dot to an oscillating electric field $\mathbf{\tilde{E}}$ and the hyperfine field of an ensemble of nuclei~\cite{footnote1}.  For a parabolic quantum dot with zero-order Hamiltonian $H_0=\hbar^2{\bf k}^2/2m+m\omega_0^2{\bf r}^2/2-|g|\mu_B({\bf S}\cdot{\bf B})$, the calculation can be simplified by performing a canonical transformation $U=\exp[i{\bf k}\cdot{\bf R}(t)]$ to a frame moving with the dot, where ${\bf R}(t)=-e{\bf \tilde{E}}(t)/m\omega_0^2$.  Here $\mathbf{S}=\mbox{\boldmath$\sigma$}/2$, $\omega_0$ is the confinement frequency, and $\hbar\mathbf{k}$ is the quasimomentum.  The transformed hyperfine Hamiltonian reads $H_{\rm hf}^U=A\Sigma_j\delta({\bf r}+{\bf R}(t)-{\bf r}_j)({\bf I}_j\cdot{\bf S})$, with $A$ the hyperfine coupling constant and the summation running over all nuclear spins $\mathbf{I}_j$.  After averaging over the orbital ground-state wave function $\psi_0({\bf r})$  and expanding in ${\bf R}(t)$ (assumed small compared to the dot size) this becomes $H_{\rm hf}^U(t)={\bf J}(t)\cdot\mbox{\boldmath$\sigma$}$, where ${\bf J}(t)$ is an operator in all ${\bf I}_j$.  Choosing the $z$-axis in spin space along $\bf B$, the components of ${\bf J}(t)$ are $J_z={1\over2}A\sum_j\psi_0^2({\bf r}_j)I_j^z$ and
\begin{equation}
 J_\pm(t)={{eA}\over{m\omega_0^2}}\sum_j\psi_0({\bf r}_j){\bf \tilde{E}}(t)\cdot{\bf \nabla}\psi_0({\bf r}_j)I_j^\pm.
\label{eq1}
\end{equation}
  The time-dependent off-diagonal components $J_\pm(t)$ drive EDSR, while the diagonal component $J_z$ describes a detuning of EDSR from the Larmor frequency $\omega_L$ by an amount $\omega_z$ randomly distributed as $\rho(\omega_z)=\exp(-\omega_z^2/\Delta^2)/(\Delta\sqrt{\pi})$~\cite{MerkulovNuclei}; time dependent corrections to $\omega_z$ are disregarded.
The dispersion $\Delta$ and the Rabi frequency $\Omega_R$ are the root-mean-square values of $J_z$ and $J_\pm$. The former is dominated by symmetric fluctuations of $\mathbf{I}_j$, the latter by antisymmetric ones because $\mathbf{\tilde{E}}\cdot\nabla\psi_0({\bf r})$ is odd with respect to the $\mathbf{\tilde{E}}$ projection of ${\bf r}$. Finally,
\begin{equation}
\Delta={{A}\over{2\hbar}}\sqrt{{{I(I+1)m\omega_0n_0}\over{3\pi\hbar d}}},\,
\Omega_R={{e\tilde{E}A}\over{\hbar^2\omega_0}}\sqrt{{{I(I+1)n_0}\over{32\pi d}}}\,,
\label{eq2}
\end{equation}
with $I=3/2$, $n_0$ the nuclear concentration, and $d$ the vertical confinement.  Remarkably, $\Omega_R$ is independent of $B$; this is in contrast to spin-orbit-mediated EDSR of localized electrons, where Kramers' theorem requires that the Rabi frequency vanish linearly as $B\rightarrow0$~\cite{RashbaShekaBook, LevitovRashba, GolovachLoss}.

In an instantaneous nuclear spin configuration with detuning $\delta\omega=2\pi f-(\omega_L+\omega_z)$ and Rabi frequency $\Omega$, the spin-flip probability from an initial $\uparrow$ spin state is~\cite{RabiOscillations}:
\begin{equation}
p_\downarrow(\tau_\mathrm{EDSR})={{\Omega^2}\over{\left(\delta\omega/2\right)^2+\Omega^2}}
\sin^2{\left[\sqrt{\left(\delta\omega/2\right)^2+\Omega^2}~\tau_\mathrm{EDSR}\right]}\,.
\label{eq3}
\end{equation}
(We neglect the comparatively slow relaxation and dephasing of the electron spin~\cite{PettaT2}.) To compare with the time-averaged data of Fig.~3, we average Eq.~\ref{eq3} over $\omega_z$ with weight $\rho(\omega_z)$ and over $\Omega$ with weight $\rho(\Omega)=2\Omega\exp(-\Omega^2/\Omega_R^2)/\Omega_R^2$. This latter distribution arises because the $J_\pm$ acquire Gaussian-distributed contributions from both $I^x_j$ and $I^y_j$ components of the nuclear spins.  The resulting spin-flip probability $\overline {p}_\downarrow (\tau_\mathrm{EDSR};\Delta, \Omega_R)$ shows only a remnant of Rabi oscillations as a weak overshoot at $\tau_\mathrm{EDSR}\sim\Omega_R^{-1}$.  Absence of Rabi oscillations is a specific property of hyperfine-driven EDSR originating because $J_\pm$ average to zero.

To compare with data, the probability $\overline {p}_\downarrow(\tau_\mathrm{EDSR})$ is scaled by a QPC sensitivity $V^0_\mathrm{QPC}$ to convert to a voltage $\delta V_\mathrm{QPC}^\mathrm{peak}$. The Larmor frequency spread, $\Delta=2\pi \times 28$~MHz, is taken as the quadrature sum of the jitter amplitude seen in Fig.~2 and half the frequency modulation depth. $\Omega_R$ and $V^0_\mathrm{QPC}$ are parameters in numerical fits: The 44~mT data (green curve in Fig.~3(a)) give $\Omega_R = 1.7\times10^6$ $\mathrm{s}^{-1}$ and $V^0_\mathrm{QPC}=2.4$~$\mu$V.  Holding $V^0_\mathrm{QPC}$ to this value, the 550~mT data give $\Omega_R=1.8\times10^6$~$\mathrm{s}^{-1}$ (blue curve in Fig.~3(a)) and the 185~mT data give values of $\Omega_R$ in Fig.~3(b).  Resulting $\Omega_R$ values increase as $\sqrt{P_\mathrm{MW}}$ (Fig.~3(b)) and are independent of $B$, both consistent with Eq.~1.  The $B$-independence of $\Omega_R$---also evident in the EDSR intensity in Fig.~2---and the absence of Rabi oscillations support our interpretation of hyperfine-mediated EDSR in the parameter range investigated.

Estimating $\hbar\omega_0 \sim 1$ meV~\cite{HansonZeeman}, $\tilde{E}\sim 6\times10^3$ Vm$^{-1}$ at maximum applied power~\cite{footnote2}, $d\sim5$ nm, and using the known values $n_0=4\times 10^{28}$ m$^{-3}$ and $An_0$=90 $\mu$eV \cite{PagetHyperfine} we calculate $\Omega_R\sim11\times10^6$ $\mathrm{s}^{-1}$, an order of magnitude larger than measured.  The discrepancy may reflect uncertainty in our estimate of $\tilde{E}$, or its spatial inhomogeneity.

Above, we generalized a mean-field description of the hyperfine interaction~\cite{KhaetskiiNuclei, MerkulovNuclei} to the resonance regime, where flip-flop processes make its applicability not obvious.  We speculate the weak overshoot in the theoretical curves in Fig.~3, which is not seen in the data, is washed out by the inhomogeneous (rate $\propto \psi_0^2$) precession of nuclear spins in the hyperfine field.

 \begin{figure}[!t]
\center \label{fig:fig4}
\includegraphics[width=3in]{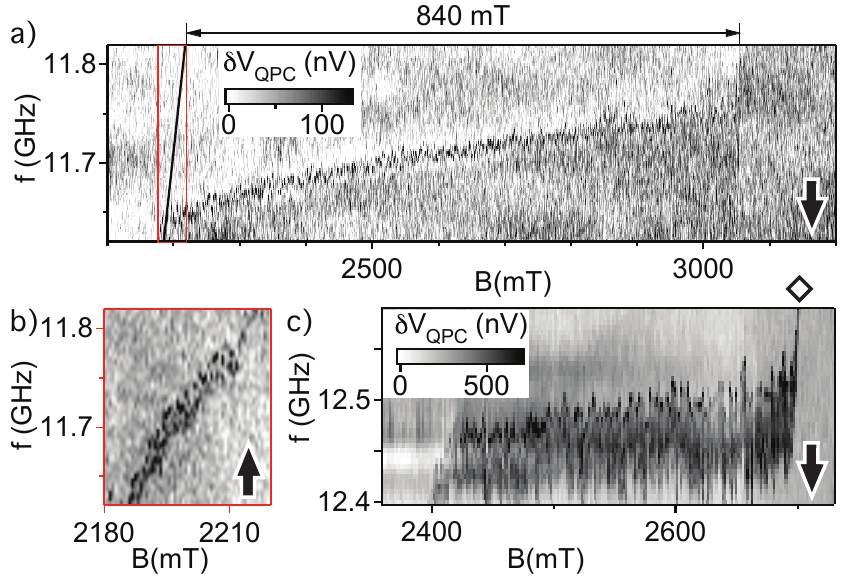}
\vspace{-0.3 cm} \caption{\footnotesize{(Color online) (a) Evolution of the resonance position as $B$ is swept upwards under polarizing conditions.  Nuclear polarization partly counteracts $B$, moving the resonance away from its equilibrium position (black diagonal line) by up to 840 mT.  (b) Similar data taken at lower power and opposite frequency sweep direction, showing approximately the equilibrium resonance position. (Gray scale as in (a)) (c) Similar data as in (a), with faster sweep rate, showing more clearly the displacement and subsequent return to equilibrium of the resonance.  $\diamondsuit$ marks the escape of the resonance from the swept frequency window.  In all plots, arrows denote frequency sweep direction.}}
\end{figure}

Consistent with a hyperfine mechanism, this EDSR effect can create nuclear polarization~\cite{BaughPolarization}.  If $f$ is scanned repeatedly over the resonance at high power, a shift of the resonance develops over $\sim$100~s (not shown), corresponding to a nuclear spin alignment parallel to $\mathbf{B}$.  The effect is stronger at higher $B$.  In Fig.~4(a), we show how to build up a substantial polarization:  While slowly increasing $B$, we scan $f$ repeatedly downwards, i.\,e., in the direction which approximately tracks the moving resonance. From the maximum line displacement from equilibrium, an effective hyperfine field of 840~mT can be read off, corresponding to a nuclear polarization of $\sim 16\%$.  Figure 4(b) shows similar data for lower power and opposite frequency sweep direction, indicating the approximate equilibrium line position.  Fig.~4(c), similar to Fig.~4(b) but with a faster sweep rate, makes the displacement and eventual escape of the resonance clearer although the maximum polarization is less.

The resonance shift is observed to be towards lower frequency, corresponding to a nuclear polarization parallel to $\mathbf{B}$.  This may be understood if the pulse cycle preferentially prepares the ground state $T_+$ over $T_-$, either because it is more efficiently loaded or because of spin-orbit-mediated relaxation.  EDSR then transfers this electron polarization to the nuclei.  We note that the line shift is opposite to what is given by the the usual Overhauser mechanism for inducing nuclear polarization via electron resonance~\cite{OverhauserPolarization, GueronRyter}.

\begin{figure}[!t]
\center \label{fig:fig5}
\includegraphics{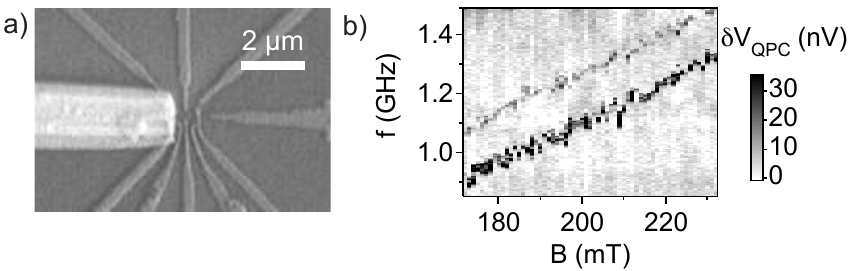}
\vspace{-0.3 cm} \caption{\footnotesize{(Color online) (a)  A device similar to that of Fig.~1, incorporating a micromagnet.  (b) The associated split EDSR line.}}
\end{figure}

In quantum information applications, it is desirable to address individual spins selectively~\cite{LossDiVincenzo}.  A scheme to allow this is presented in Fig.~5.  An otherwise similar device~(Fig. 5(a)) incorporated a 100~nm thick micron-scale permalloy (84\% Ni, 16\% Fe) magnet over 35~nm of atomic-layer-deposited alumina~\cite{TokuraSlanting, PioroLadriere}.  In this device, measured with $\mathbf{B}$ normal to the heterostructure plane, the EDSR line was frequently split by $10-20$~mT (Fig.~5(b).)  This splitting, not observed without the magnet, is considerably larger than the Overhauser field fluctuations and presumably reflects a magnetic field difference between the dots. With separated resonance lines for right and left dots, different spins can be addressed by matching $f$ to the local resonance condition.

We acknowledge useful discussions with Al.~L.~Efros, H.-A.~Engel, F.~H.~L.~Koppens, J.~R.~Petta, D.~J.~Reilly, M.~S.~Rudner, and J.~M.~Taylor. We acknowledge support from the DTO and from DARPA.

\small

\begin{thebibliography}{34}
\bibitem{footnote2} Microwave power at each frequency is calibrated using the non-resonant lifting of spin blockade, which we interpret to as where the microwave amplitude exceeds the gate voltage from C to the nearest charge transition.  The data in Fig.~3(a) and the last data point in Fig.~3(b) use power $2\pm1$~dB below threshold, corresponding to 3.2~mV.  Dropped uniformly across the whole device this amplitude gives a field $\tilde{E}\sim6\times10^3$~Vm$^{-1}$.  Sourced power in Fig.~3(a) was $P_\mathrm{MW}=0.6$ mW. The 41~dB attenuation to the sample holder then gives good agreement with calibrated value at those values of $\overline{f}$.
\end{thebibliography}

\end{document}